\documentclass[final,1p,times,twocolumn]{elsarticle}

\usepackage{amssymb}

 \biboptions{compress}

\begin{document}

\begin{frontmatter}

\title{Geometric measure of quantum discord over two-sided projective measurements}

 \author{Jianwei Xu\corref{cor1}}

 \cortext[cor1]{Corresponding author. Tel.: +86 28 85412379; fax: +86 28 8541 0252}
\ead{xxujianwei@yahoo.cn}

\address{Key Laboratory for Radiation Physics and Technology, Institute
of Nuclear Science and Technology, Sichuan University, Chengdu 610065, China}

\begin{abstract}
The original definition of quantum discord of bipartite states was defined over one-sided projective measurements, it describes quantum correlation more extensively than entanglement. Dakic, Vedral, and Brukner [Phys. Rev. Lett. 105 (2010) 190502] introduced a geometric measure for this quantum discord, and Luo, Fu [Phys. Rev. A 82 (2010) 034302] simplified the variation expression of it. In this paper we introduce a geometric measure for the quantum discord over two-sided projective measurements. A simplified expression and a lower bound of this geometric measure are derived and explicit expressions are obtained for some special cases.
\end{abstract}


\end{frontmatter}

\section{Introduction}

Quantum entanglement is by far the most famous and best studied kind of quantum correlation, and leads to powerful applications \cite{Nielson2000,Horodecki2009}. While interest remains strong, recent researches have explored another quantum correlation other than entanglement, called quantum discord \cite{Ollivier2001,Henderson2001}, which may be employed as alternative resources for quantum technology \cite {Datta2008,Lanyon2008,Datta2009}.  From theoretic points of view, operational interpretations of quantum discord have been proposed, the links between quantum discord with other concepts have been discussed,  such as Maxwell's demon \cite{Zurek2003,Brodutch2010}, completely positive maps \cite{Shabani2009}, and relative entropy \cite{Modi2010}. At the same time, the characteristics of quantum
discord in some physical models and in information processing have been studied \cite{Werlang2009,Werlang2010,Wang2010,Soares2010,Altintas2010,Li2011}.

But the awkward situation is, till now the analytical expressions of quantum discord are found only for few special states \cite{Luo2008,Ali2010,Giorda2010,Bylicka2010,Ferraro2010}. The problem arises from the variation expression of original definition of quantum discord. Analytical expression is very useful for investigating the dynamics in physical systems \cite{Werlang2009,Yu2004,Altintas2010-2}. Very recently, Dakic, Vedral, and Brukner introduced \cite{Dakic2010} a geometric measure for quantum discord. As one of the most striking results of this measure, they obtained \cite{Dakic2010} the analytical expression for any two qubits states. Also,  Luo and Fu \cite{Luo2010} simplified the expression of this geometric measure, and derived a lower bound for any quantum state.

In the same spirit, in this paper we introduce a geometric measure for the quantum discord over two-sided projective measurements (Sec.3). We also simplify the expression and provide a lower bound for this geometric measure (Sec.4). As examples, we derive some explicit expressions of this geometric measure for some special quantum states (Sec.5).

\section{Geometric measure of quantum discord over one-sided projective measurements}
For clarity, we first give some definitions about quantum entropy, conditional entropy, mutual information and projective measurement. Let $H^A,H^B$ be the Hilbert spaces of quantum
systems $A$, $B$, respectively, with $dimH^A=n_A$, $dimH^B=n_B$. $I_A,I_B,I_{AB}$
are the identity operators on $H^A,H^B$ and $H^A\otimes H^B$. The reduced
density matrices of a state $\rho ^{AB}$ on $H^A\otimes H^B$ are $\rho ^A=tr_B\rho^{AB} $, $\rho ^B=tr_A\rho^{AB}.$ For density operators $\rho ,\sigma $
on a Hilbert space H, the entropy of $\rho$ is defined as $S(\rho )=-tr(\rho \log \rho
)$ $(\log \rho =\log _2\rho )$, the conditional entropy of $\rho ^{AB}$ (with respect to A) is defined as $S(\rho ^{AB})-S(\rho ^A)$. The mutual information of $\rho ^{AB} $ is defined as
$S(\rho ^A)+S(\rho ^B)-S(\rho ^{AB})$, which is nonnegative, and vanishing only when $\rho ^{AB}=\rho ^A\otimes
\rho ^B$ (\cite{Nielson2000}, 11.3.4). A general measurement on $\rho ^{AB}$
is denoted by a set of operators $\Phi =\{\Phi _\alpha \}_\alpha $
satisfying $\sum_\alpha \Phi _\alpha ^{\dagger } \Phi _\alpha =I_{AB}$, here $%
\dagger $ denotes Hermitian adjoint, and $\{\Phi _\alpha \}_\alpha $ performs $%
\rho ^{AB}$ as $\widetilde{\rho ^{AB}}=\sum_\alpha \Phi _\alpha \rho
^{AB}\Phi _\alpha ^{\dagger }.$ When $\Phi _\alpha =\Pi_\alpha \otimes I_B$, where
$\Pi _\alpha =|\alpha \rangle \langle \alpha |$ and $\{|\alpha \rangle \}_{\alpha =1}^{n_A}$ is an orthonormal basis for $H^A,$ we
call $\{\Pi _\alpha \otimes I_B\}_\alpha $ a one-sided projective
measurement. We call $\{\Pi _{\alpha \beta }\}_{\alpha \beta }$ a
two-sided projective measurements, where $\Pi _{\alpha \beta }=|\alpha
\rangle \langle \alpha |\otimes |\beta \rangle \langle \beta |$, and $\{|\beta \rangle \}_{\beta =1}^{n_B}$ is an orthonormal basis of $H^B$. For
simplicity, we sometimes simply write $\Pi_\alpha \otimes I_B$ as $\Pi_\alpha $. We use $\widetilde{\rho ^{AB}}$ to denote
the state its initial state is $\rho ^{AB}$ and experienced a
measurement, and $\widetilde{\rho ^A}=tr_B\widetilde{\rho ^{AB}},\widetilde{%
\rho ^B}=tr_A\widetilde{\rho ^{AB}}$.

Now recall that the original definition of quantum discord of $\rho ^{AB}$
was defined over one-sided projective measurements (with respect to A) as \cite
{Ollivier2001}
\begin{equation}
D_A(\rho ^{AB})=S(\rho ^A)-S(\rho ^{AB})+\inf_{\{\Pi _\alpha \otimes
I_B\}_\alpha }[\Sigma_{\alpha}p_\alpha S(\widetilde{\rho _\alpha ^B}/p_\alpha )],
\end{equation}
where inf is taken over all projective measurements on A, $\widetilde{%
\rho _\alpha ^B}=tr_A(\Pi _\alpha \rho ^{AB}\Pi _\alpha )$, $p_\alpha =tr_B%
\widetilde{\rho _\alpha ^B}$. $D_B(\rho ^{AB})$ can be defined similarly.

By the joint entropy theorem (\cite{Nielson2000}, 11.3.2), Eq.(1) can be rewritten as \cite{Ollivier2001}
\begin{equation}
D_A(\rho ^{AB})=S(\rho ^A)-S(\rho ^{AB})+\inf_{\{\Pi _\alpha \otimes
I_B\}_\alpha }[S(\widetilde{\rho ^{AB}})-S(\widetilde{\rho ^A})].
\end{equation}

A state $\rho ^{AB}$ satisfying $D_A(\rho ^{AB})=0$ is called a classical state, it
can be proved that \cite{Ollivier2001}
\begin{eqnarray}
D_A(\rho ^{AB})&\geq& 0,\\
D_A(\rho ^{AB})&=&0\Longleftrightarrow \rho ^{AB}=\sum_{\alpha
=1}^{n_A}p_\alpha |\alpha \rangle \langle \alpha |\otimes \rho _\alpha ^B,
\end{eqnarray}
where, $\left\{ |\alpha \rangle \right\} _{\alpha =1}^{n_A}$ is an arbitrary orthonormal set of $H^A$, $p_\alpha \geq 0$,  $\sum_{\alpha =1}^{n_A}p_\alpha =1$, $\rho _\alpha ^B$ are density operators on $H^B$.

Although the set of all states $\rho ^{AB}$ satisfying $D_A(\rho ^{AB})=0$
is not a convex set, a technical definition of geometric measure of quantum
discord of $\rho ^{AB}$ over one-sided projective measurements on A can be
defined as
\begin{equation}
D_A^G(\rho ^{AB})=\inf_{\sigma ^{AB}} d(\rho ^{AB},\sigma ^{AB}),
\end{equation}
where d is a distance defined on density operators on $H^A\otimes H^B,$ and
inf runs over all $\sigma ^{AB}$ with $D_A^{P}(\sigma ^{AB})=0$. $%
D_B^G(\rho ^{AB})$ can be defined similarly. One of such geometric measure is as follows.

Let $L(H^A)$ be the real linear space of all Hermitian operators on $H^A$
, and define the inner product $\langle X|X'\rangle =tr_A(XX')$ \smallskip for
any $X,X'\in L(H^A)$, then $L(H^A)$ becomes a real Hilbert space with
dimension $n_A^2$. The Hilbert spaces $L(H^B)$ and $L(H^A\otimes H^B)$ are
defined similarly. A geometric measure of quantum discord of $\rho ^{AB}$
over one-sided projective measurements on A can then be defined as \cite
{Dakic2010}
\begin{equation}
D_A^G(\rho ^{AB})=\inf_{\sigma ^{AB}}||\rho ^{AB}-\sigma ^{AB}||^2,
\end{equation}
where $||\rho ^{AB}-\sigma ^{AB}||^2=tr[(\rho ^{AB}-\sigma ^{AB})^2]$, inf
takes all $\sigma ^{AB}$ that $D_A^P(\sigma ^{AB})=0.$ Some analytical
solutions of $D_A^G(\rho ^{AB})$ were obtained \cite{Dakic2010}. Moreover,
it has been shown that Eq.(6) can be simplified as \cite{Luo2010}
\begin{equation}
D_A^G(\rho ^{AB})=\inf_{\{\Pi _\alpha \}_{\alpha}}||\rho ^{AB}-\sum_\alpha \Pi
_\alpha \rho ^{AB}\Pi _\alpha ||^2.
\end{equation}

\section{Geometric measure of quantum discord over two-sided projective measurements}

In this section, we propose a geometric measure under two-sided projective
measurements.

 The original definition of quantum discord over one-sided
projective measurements in Eq.(1) or Eq.(2) has the intuitive physical meaning that $D_A(\rho ^{AB})$ is the minimal loss of mutual information or conditional entropy due to all one-sided projective measurements. A direct way to define the quantum discord over two-sided projective measurements then is \cite{Ollivier2001, Maziero2010}
\begin{eqnarray}
D_{AB}(\rho ^{AB})=S(\rho ^A)+S(\rho ^B)-S(\rho ^{AB})
+\inf_{\{\Pi _{\alpha \beta }\}_{\alpha \beta }}[S(\widetilde{\rho ^{AB}}
)-S(\widetilde{\rho ^A})-S(\widetilde{\rho ^B})].
\end{eqnarray}
Where, inf takes all two-sided projective measurements. By the experiences of optimization about Eq.(1) or Eq.(2), it seems that Eq.(8) will be very difficult to optimize excepting some very special states. So, we introduce a geometric measure of it, just as what have done in the one-sided case \cite{Dakic2010, Luo2010}. To do so, we first prove that $D_{AB}(\rho^{AB})$ in Eq.(8) is nonnegative for any state (then $D_{AB}(\rho ^{AB})$ is a valid measure), and next we need to find the set of all states  $\rho ^{AB}$ that $D_{AB}(\rho ^{AB})=0$. This is Theorem 1 below.

\emph{Theorem 1}. It holds that
\begin{eqnarray}
D_{AB}(\rho ^{AB}) &\geq &0, \\
D_{AB}(\rho ^{AB}) &=&0\Longleftrightarrow \rho ^{AB}=\sum_{\alpha \beta
}p_{\alpha \beta }\Pi _{\alpha \beta },
\end{eqnarray}
where $\{\Pi _{\alpha \beta }\}_{\alpha \beta }$ is an arbitrary two-sided projective measurement, $\{p_{\alpha \beta }\}_{\alpha \beta }$ is an arbitrary probability distrbution, that is  $p_{\alpha \beta }\geq 0$, $\sum_{\alpha \beta }p_{\alpha \beta }=1$.

\emph{proof.} Given a two-sided projective measurement $\{\Pi _{\alpha \beta }\}_{\alpha \beta }$, notice that
\begin{eqnarray}
\widetilde{\rho ^{AB}}=\sum_{\alpha \beta }\Pi _{\alpha \beta }\rho ^{AB}\Pi _{\alpha \beta }=\sum_\beta I_A\otimes \Pi _\beta (\overline{\rho _1^{AB}})I_A\otimes \Pi _\beta ,
\end{eqnarray}
where
\begin{eqnarray}
\overline{\rho _1^{AB}}=\sum_\alpha \Pi _\alpha \otimes I_B\rho ^{AB}\Pi _\alpha \otimes I_B.
\end{eqnarray}	
We expand $\rho ^{AB}$ ,  $\overline{\rho _1^{AB}}$ ,  $\widetilde{\rho ^{AB}}$ and their reduced density operators in the bases $\{|\alpha \rangle \}_{\alpha =1}^{n_A}=\{|\alpha ^{\prime }\rangle \}_{\alpha ^{\prime }=1}^{n_A}$ and $\{|\beta \rangle \}_{\beta =1}^{n_B}=\{|\beta ^{\prime }\rangle \}_{\beta ^{\prime }=1}^{n_B}$ as
	
$$\rho ^{AB}=\sum_{\alpha \alpha ^{\prime }\beta \beta ^{\prime }}\rho _{\alpha \alpha ^{\prime }\beta \beta ^{\prime }}^{AB}|\alpha \rangle \langle \alpha ^{\prime }|\otimes |\beta \rangle \langle \beta ^{\prime }|,\eqno (13.1)$$
	
$$\rho ^A=\sum_{\alpha \alpha ^{\prime }\beta }\rho _{\alpha \alpha ^{\prime }\beta \beta }^{AB}|\alpha \rangle \langle \alpha ^{\prime }|,\eqno (13.2)$$
	
$$\ \rho ^B=\sum_{\alpha \beta \beta ^{\prime }}\rho _{\alpha \alpha \beta \beta ^{\prime }}^{AB}|\beta \rangle \langle \beta ^{\prime }|,\eqno (13.3)$$
	
$$\overline{\rho _1^{AB}}=\sum_\alpha \Pi _\alpha \rho ^{AB}\Pi _\alpha =\sum_{\alpha \beta \beta ^{\prime }}\rho _{\alpha \alpha \beta \beta ^{\prime }}^{AB}|\alpha \rangle \langle \alpha |\otimes |\beta \rangle \langle \beta ^{\prime }|,\eqno (14.1)$$
	
$$\overline{\rho _1^A}=tr_B\overline{\rho _1^{AB}}=\sum_{\alpha \beta }\rho _{\alpha \alpha \beta \beta }^{AB}|\alpha \rangle \langle \alpha |,\eqno (14.2)$$
	
$$\overline{\rho _1^B}=tr_A\overline{\rho _1^{AB}}=\sum_{\alpha \beta \beta ^{\prime }}\rho _{\alpha \alpha \beta \beta ^{\prime }}^{AB}|\beta \rangle \langle \beta ^{\prime }|,\eqno (14.3)$$
$$\widetilde{\rho ^{AB}}=\sum_\beta \Pi _\beta (\overline{\rho _1^{AB}})\Pi _\beta =\sum_{\alpha \beta }\rho _{\alpha \alpha \beta \beta }^{AB}|\alpha \rangle \langle \alpha |\otimes |\beta \rangle \langle \beta |,\eqno (15.1)$$
	
$$\widetilde{\rho ^A}=tr_B\widetilde{\rho ^{AB}}=\sum_{\alpha \beta }\rho _{\alpha \alpha \beta \beta }^{AB}|\alpha \rangle \langle \alpha |,\ \ \ \eqno (15.2)$$
	
$$\ \widetilde{\rho ^B}=tr_A\widetilde{\rho ^{AB}}=\sum_{\alpha \beta }\rho _{\alpha \alpha \beta \beta }^{AB}|\beta \rangle \langle \beta |.\eqno (15.3)$$
\setcounter {equation}{15}
Where $\rho _{\alpha \alpha ^{\prime }\beta \beta ^{\prime }}^{AB}=\langle \alpha \beta |\rho ^{AB}|\alpha ^{\prime }\beta \rangle$.
From Eq.(13.3) and Eq.(14.3), Eq.(14.2) and Eq.(15.2),  we have
\begin{eqnarray}
\rho ^B=\overline{\rho _1^B}, \ \ \ \overline{\rho _1^A}=\widetilde{\rho ^A}.
\end{eqnarray}
Then
\begin{eqnarray}
&&[S(\widetilde{\rho ^{AB}})-S(\widetilde{\rho ^A})-S(\widetilde{\rho ^B})]-[S(\rho ^{AB})-S(\rho ^A)-S(\rho ^B))]	\nonumber \\
&=&\{[S(\widetilde{\rho ^{AB}})-S(\widetilde{\rho ^A})-S(\widetilde{\rho ^B})]-[S(\overline{\rho _1^{AB}})-S(\overline{\rho _1^A})-S(\overline{\rho _1^B})]\}	\nonumber \\
&&+\{[S(\overline{\rho _1^{AB}})-S(\overline{\rho _1^A})-S(\overline{\rho _1^B})]-[S(\rho ^{AB})-S(\rho ^A)-S(\rho ^B))]\}	\nonumber \\
&=&\{[S(\widetilde{\rho ^{AB}})-S(\widetilde{\rho ^B})]-[S(\overline{\rho _1^{AB}})-S(\overline{\rho _1^B})]\}	
+\{[S(\overline{\rho _1^{AB}})-S(\overline{\rho _1^A})]-[S(\rho ^{AB})-S(\rho ^A)]\}.
\end{eqnarray}
	
From Eq.(2) and Eq.(3), it can be seen that the two expressions in the two curly braces of last line in Eq.(17) are both nonnegative, then we obtain Eq.(9).
	
To prove Eq.(10), suppose $D_{AB}(\rho ^{AB})=0$ and the zero can be achieved by the two-sided projective measurement $\{\Pi _{\alpha \beta }\}_{\alpha \beta }$. Again from Eq.(17), it follows that the two expressions in the two curly braces of last line in Eq.(17) are both vanishing. Then $D_A(\rho ^{AB})=0$ and $D_B(\overline{\rho ^{AB}})=0$. Similarly, when we repeat the above program substituting  $\overline{\rho _2^{AB}}$ by $\overline{\rho _2^{AB}}=\sum_\beta I_A\otimes \Pi _\beta (\rho ^{AB})I_A\otimes \Pi _\beta$, 	
we will obtain $D_B(\rho ^{AB})=0$. Combining $D_A(\rho ^{AB})=0$, $D_B(\rho ^{AB})=0$, and Eq.(4), we steadily obtain Eq.(10). That is to say,
\begin{eqnarray}
D_{AB}(\rho ^{AB})=0\Longleftrightarrow D_A(\rho ^{AB})=D_B(\rho ^{AB})=0.
\end{eqnarray}
We then complete this proof.

The intuitive meaning of Eq.(8) is that $D_{AB}(\rho ^{AB})$ is the
minimal loss of mutual information over all two-sided projective
measurements. From Eq.(3) and Eq,(10), or from Eq.(18), we see that $D_{AB}(\rho ^{AB})$ captures more correlation
than $D_A(\rho ^{AB})$ in the sense
\begin{equation}
D_{AB}(\rho ^{AB})=0\Rightarrow D_{A}(\rho ^{AB})=0.
\end{equation}

Similar to Eq.(6), we also define a geometric measure of quantum
discord over two-sided projective measurements as
\begin{equation}
D_{AB}^G(\rho ^{AB})=\inf_{\chi ^{AB}}||\rho ^{AB}-\chi ^{AB}||^2,
\end{equation}
where $||\rho ^{AB}-\chi ^{AB}||^2=tr[(\rho ^{AB}-\chi ^{AB})^2]$, inf takes all $\chi ^{AB}$ that $D_{AB}(\chi ^{AB})=0$. From the definitions of $D_A^G(\rho ^{AB})$ and $D_{AB}^G(\rho ^{AB})$, and
Eq. (19), it can be easily found that
\begin{equation}
D_{AB}^G(\rho ^{AB})\geq \max \{D_A^G(\rho ^{AB}),D_B^G(\rho ^{AB})\}.
\end{equation}

\section{Simplification and a lower bound of Eq.(20)}

Theorem 2 below will simplify Eq. (20).

\emph{Theorem 2.} $D_{AB}^G(\rho ^{AB})$ is defined in Eq. (20), then
\begin{eqnarray}
D_{AB}^G(\rho ^{AB})&=&\inf_{\{\Pi _{\alpha \beta }\}_{\alpha \beta
}}||\rho ^{AB}-\sum_{\alpha \beta }\Pi _{\alpha \beta }\rho ^{AB}\Pi
_{\alpha \beta }||^2   \\
&=&tr[(\rho ^{AB})^2]-\sup_{\{\Pi _{\alpha \beta }\}_{\alpha \beta
}}||\sum_{\alpha \beta }\Pi _{\alpha \beta }\rho ^{AB}\Pi _{\alpha \beta
}||^2,
\end{eqnarray}
where inf and sup take over all two-sided projective measurements $\{\Pi
_{\alpha \beta }\}_{\alpha \beta }.$

\emph{Proof}. For any $\chi ^{AB}$ that $D_{AB}(\chi ^{AB})=0$, suppose
\begin{eqnarray}
\chi ^{AB}=\sum_{\alpha \beta }p_{\alpha \beta }|\alpha \rangle \langle
\alpha |\otimes |\beta \rangle \langle \beta |.
\end{eqnarray}

Where $\{|\alpha \rangle \}_{\alpha =1}^{n_A}$, $\{|\beta \rangle \}_{\beta =1}^{n_B}$ are orthonormal bases for $H^A$ and $H^B$, $p_{\alpha \beta }\geq 0$, $\sum_{\alpha \beta }p_{\alpha \beta }=1$. We expand $\rho ^{AB}$ in the bases $\{|\alpha \rangle \}_{\alpha =1}^{n_A}=\{|\alpha ^{\prime }\rangle \}_{\alpha ^{\prime }=1}^{n_A}$ and $\{|\beta \rangle \}_{\beta =1}^{n_B}=\{|\beta ^{\prime }\rangle \}_{\beta ^{\prime }=1}^{n_B}$ as

\begin{eqnarray}
\rho ^{AB}=\sum_{\alpha \alpha ^{\prime }\beta \beta ^{\prime }}\rho
_{\alpha \alpha ^{\prime }\beta \beta ^{\prime }}^{AB}|\alpha \rangle
\langle \alpha ^{\prime }|\otimes |\beta \rangle \langle \beta ^{\prime }|.
\end{eqnarray}
where  $\rho _{\alpha \alpha ^{\prime }\beta \beta ^{\prime }}^{AB}=\langle \alpha \beta |\rho ^{AB}|\alpha ^{\prime }\beta ^{\prime }\rangle $, and $\rho _{\alpha \alpha \beta \beta }^{AB}=\langle \alpha \beta |\rho ^{AB}|\alpha \beta \rangle \geq 0$, $\sum_{\alpha \beta }\rho _{\alpha \alpha \beta \beta }^{AB}=1$. Consequently,
\begin{eqnarray}
&&||\rho ^{AB}-\chi ^{AB}||^2  \nonumber \\
&=&tr[(\rho ^{AB})^2]-2\sum_{\alpha \beta }p_{\alpha \beta }\rho _{\alpha
\alpha \beta \beta }^{AB}+\sum_{\alpha \beta }p_{\alpha \beta }^2  \nonumber
\\
&=&tr[(\rho ^{AB})^2]-\sum_{\alpha \beta }(\rho _{\alpha \alpha \beta \beta
}^{AB})^2+\sum_{\alpha \beta }(p_{\alpha \beta }-\rho _{\alpha \alpha \beta
\beta }^{AB})^2.
\end{eqnarray}
By choosing $p_{\alpha \beta }=\rho _{\alpha \alpha \beta \beta }^{AB}$,
i.e., $\chi ^{AB}=\Sigma _{\alpha \beta }\Pi _{\alpha \beta }\rho ^{AB}\Pi
_{\alpha \beta }$, we then attain Theorem 3.

We would rather like to give another expression of Theorem 2, a lower
bound of $D_{AB}^G(\rho ^{AB})$ will follow from it, that is Theorem 3 below.

\emph{Theorem 3.}  $D_{AB}^G(\rho ^{AB})$ is defined in Eq.
(20), then
\begin{eqnarray}
D_{AB}^G(\rho ^{AB}) &=&tr(CC^t)-\sup_{AB}tr(ACB^tBC^tA^t), \\
D_{AB}^G(\rho ^{AB}) &\geq&tr(CC^t)-\sum_{k=1}^{\min \{n_{A},n_{B}\}}\lambda _k.
\end{eqnarray}
Where $\lambda _k$ are the eigenvalues of $CC^t$ listed in decreasing order
(counting multiplicity), $t$ denotes transpose. Real matrices $%
A,B,C$ are specified as follows: given orthonormal bases $\{X_i\}_{i=1}^{n_A^2}$ for $%
L(H^A)$ and $\{Y_j\}_{j=1}^{n_B^2}$ for $L(H^B)$. Let $\rho
^{AB}=\sum_{ij}C_{ij}X_i\otimes Y_j$, then matrix $C=(C_{ij})$. For any
orthonormal bases $\{|\alpha \rangle \}_{\alpha =1}^{n_A}$ for $H^A$ and $\{|\beta \rangle \}_{\beta =1}^{n_B}$ for $H^B$, let $|\alpha \rangle \langle
\alpha |=\sum {}_{i=1}^{n_A^2}A_{\alpha i}X_i$, $|\beta \rangle \langle
\beta |=\sum {}_{j=1}^{n_B^2}B_{\beta j}Y_j$, then matrices $A=(A_{\alpha
i}),B=(B_{\beta j}).$

To prove Eq. (27), note that $A_{\alpha i}=tr(X_i|\alpha \rangle \langle
\alpha |)=\langle \alpha |X_i|\alpha \rangle $, $B_{\beta j}=tr(Y_j|\beta
\rangle \langle \beta |)=\langle \beta |Y_j|\beta \rangle $, thus
\begin{eqnarray}
&&D_{AB}^G(\rho ^{AB})=tr[(\rho ^{AB})^2]-\sup_{\{\Pi _{\alpha \beta
}\}}||\sum_{\alpha\beta}\Pi _{\alpha \beta }\rho ^{AB}\Pi _{\alpha \beta
}||^2  \nonumber \\
&&=\sum_{ij}C_{ij}^2-\sup_{\{\Pi _{\alpha \beta }\}}||\sum_{ij\alpha \beta
}C_{ij}\langle \alpha |X_i|\alpha \rangle \langle \beta |Y_j|\beta \rangle
|\alpha \rangle \langle \alpha |\otimes |\beta \rangle \langle \beta |||^2
\nonumber \\
&&=\sum_{ij}C_{ij}^2-\sup_{AB}\sum_{\alpha \beta }(\sum_{ij}A_{\alpha
i}C_{ij}B_{\beta j})^2  \nonumber \\
&&=tr(CC^t)-\sup_{AB}tr(ACB^tBC^tA^t).  \nonumber
\end{eqnarray}

A brief proof of inequality (28) is: since \cite{Luo2010}
\begin{eqnarray}
D_A^G(\rho ^{AB}) \geq tr(CC^t)-\sum_{k=1}^{n_A}\lambda _k,  \\
D_B^G(\rho ^{AB}) \geq tr(CC^t)-\sum_{k=1}^{n_{B}}\lambda _k,
\end{eqnarray}
together with Eq.(21), so inequality (28) is surely true.

\section{Examples}
Let us consider some examples which allow explicit results for $D_{AB}^G(\rho ^{AB})$.

\emph{Example 1}. For the $m\times m$ Werner state
\begin{eqnarray}
\rho ^{AB}=\frac{m-x}{m^3-m}I_{AB}+\frac{mx-1}{m^3-m}F,\ \ x\in [-1,1],
\end{eqnarray}
with $F=\sum_{kl}|k\rangle \langle l|\otimes |l\rangle \langle k|$, $\{|k\rangle \}=\{|l\rangle \}$ is an orthonormal basis for $H^A$ $(H^A=H^B)$. Note that $F^2=I_{AB}$, $trF=m$. For any two-sided projective measurement $\{\Pi _{\alpha \beta }\}_{\alpha \beta }$,
\begin{eqnarray}
&&\sum_{\alpha \beta }\Pi _{\alpha \beta }\rho ^{AB}\Pi _{\alpha \beta }
\nonumber \\
&=&\frac{m-x}{m^3-m}I_{AB}+\frac{mx-1}{m^3-m}\sum_{\alpha \beta kl}\langle \alpha
|k\rangle \langle l|\alpha \rangle \langle \beta |l\rangle \langle k|\beta
\rangle \Pi _{\alpha \beta }  \nonumber \\
&=&\frac{m-x}{m^3-m}I_{AB}+\frac{mx-1}{m^3-m}\sum_{\alpha \beta }|\langle \alpha
|\beta \rangle |^2\Pi _{\alpha \beta }.  \nonumber
\end{eqnarray}
By Eq. (22), and applying the Lagrangian multipliers method, we get
\begin{eqnarray}
D_{AB}^G(\rho ^{AB})=\frac{(mx-1)^2}{m(m-1)(m+1)^2}.
\end{eqnarray}
That is $D_{AB}^G(\rho ^{AB})=D_A^G(\rho ^{AB})$ \cite{Luo2010}.

A Werner state is separable if and only if $x\in [0,1]$ \cite{Horodecki2009}, but $D_A^G(\rho ^{AB})=D_{AB}^G(\rho ^{AB})=0$ if and only if $x=1/m$, i.e., it is the completely mixed state.

\emph{Example 2}. For the $m\times m$ isotropic state
\begin{eqnarray}
\rho ^{AB}=\frac{1-x}{m^2-1}I_{AB}+\frac{m^2x-1}{m^2-1}M,\ \ x\in [0,1],
\end{eqnarray}
with $M=\frac 1m\sum_{kl}|k\rangle \langle l|\otimes |k\rangle \langle l|$, $\{|k\rangle \}=\{|l\rangle \}$ is an orthonormal basis for $H^A$ $(H^A=H^B)$.
Note that $M^2=M$, and $trM=1$. For any two-sided projective measurement $\{\Pi _{\alpha \beta }\}_{\alpha \beta }$
\begin{eqnarray}
&&\sum_{\alpha \beta }\Pi _{\alpha \beta }\rho ^{AB}\Pi _{\alpha \beta }
\nonumber \\
&=&\frac{1-x}{m^2-1}I_{AB}+\frac{m^2x-1}{m^2-1}\frac 1m\sum_{\alpha \beta
kl}\langle \alpha |k\rangle \langle l|\alpha \rangle \langle \beta|k\rangle
\langle l|\beta \rangle \Pi _{\alpha \beta }  \nonumber \\
&=&\frac{1-x}{m^2-1}I_{AB}+\frac{m^2x-1}{m^2-1}\frac 1m\sum_{\alpha \beta }|\langle
\alpha |\beta ^{\prime }\rangle |^2\Pi _{\alpha \beta }.  \nonumber
\end{eqnarray}
here $|\beta ^{\prime }\rangle =|\beta \rangle ^{*}$ is the complex
conjugate of $|\beta \rangle$ in the basis $\{|k\rangle \}=\{|l\rangle \}$, namely, $\langle \beta |k\rangle =\langle k|\beta ^{\prime }\rangle $, $\langle l|\beta \rangle =\langle \beta ^{\prime }|l\rangle $. Using the similar techniques in example 1
we get
\begin{eqnarray}
D_{AB}^G(\rho ^{AB})=\frac{(m^2x-1)^2}{m(m-1)(m+1)^2}.
\end{eqnarray}
That is $D_{AB}^G(\rho ^{AB})=D_A^G(\rho ^{AB})$ \cite{Luo2010}.

Recall that an isotropic state is separable if and only if $x\in [0,1/m]$ \cite{Horodecki2009},
but $D_A^G(\rho ^{AB})=D_{AB}^G(\rho ^{AB})=0$ if and only if $%
x=1/m^2,$ i.e., it is the completely mixed state.

\emph{\ Example 3}. For any two-qubit state
\begin{eqnarray}
\rho &=&\frac 14(I_{AB}+\sum_{i=1}^3x_i\sigma _i\otimes
I_B+\sum_{j=1}^3y_jI_A\otimes \sigma _j+\sum_{i,j=1}^3T_{ij}\sigma _i\otimes \sigma _j)  \nonumber \\
&=&\frac 12(X_0\otimes Y_0+\sum_{i=1}^3x_iX_i\otimes
Y_0+\sum_{j=1}^3y_jX_0\otimes Y_j+\sum_{i,j=1}^3T_{ij}X_i\otimes Y_j).
\end{eqnarray}
Where $\textbf{x}=(x_1,x_2,x_3)$, $\textbf{y}=(y_1,y_2,y_3)$ are two real vectors, $\{\sigma _i\}$ are the Pauli matrices, $\{X_0,X_1,X_2,X_3\}=\{I_A,\sigma _1,\sigma _2,\sigma _3\}/\sqrt{2}$, $\{Y_0,Y_1,Y_2,Y_3\}=\{I_B,\sigma _1,\sigma _2,\sigma _3\}/\sqrt{2}$. Note that $tr\sigma _i=0$ and $tr(\sigma _i\sigma _j)=2\delta _{ij}$, hence $\{X_0,X_1,X_2,X_3\}$ is an orthonormal basis for $L(H^A)$ , and $\{Y_0,Y_1,Y_2,Y_3\}$ is an orthonormal basis for $L(H^B)$. For any orthonormal basis $\{|\alpha \rangle \}_{\alpha =1}^2$ of $H^A$, $|\alpha \rangle \langle \alpha |\in L(H^A)$, we can write $|\alpha \rangle \langle \alpha |$ as
$$|1_A\rangle \langle 1_A|=(X_0+a_1X_1+a_2X_2+a_3X_3)/\sqrt{2}, \eqno (36.1)$$
$$|2_A\rangle \langle 2_A|=(X_0-a_1X_1-a_2X_2-a_3X_3)/\sqrt{2}.\eqno (36.2)$$
Here, $\mathbf{a}=(a_1,a_2,a_3)$ is a real vector with $\|\mathbf{a}\|^{2}=\sum_{i=1}^3a_i^2=1$. Similarly, for any orthonormal basis $\{|\beta
\rangle \}_{\alpha =1}^2$ of $H^B$, $|\beta \rangle \langle \beta |\in
L(H^B) $, we write $|\beta \rangle \langle \beta |$ as
$$|1_B\rangle \langle 1_B|=(Y_0+b_1Y_1+b_2Y_2+b_3Y_3)/\sqrt{2},\eqno (37.1)$$
$$|2_B\rangle \langle 2_B|=(Y_0-b_1Y_1-b_2Y_2-b_3Y_3)/\sqrt{2}.\eqno (37.2)$$
\setcounter {equation}{37}
Here, $\mathbf{b}=(b_1,b_2,b_3)$ is a real vector with $\|\mathbf{b}\|^{2}=\sum_{i=1}^3b_i^2=1$.
Thus, from Eq. (27), direct calculation shows that
\begin{eqnarray}
D_{AB}^G(\rho ^{AB})=\frac 14[||\mathbf{x}||^2+||\mathbf{y}||^2+tr(TT^t)]
-\frac 14\sup_{\mathbf{ab}}[(\mathbf{a}\cdot \mathbf{x})^2+(\mathbf{b}\cdot
\mathbf{y})^2+(\mathbf{a}T\mathbf{b}^t)^2].
\end{eqnarray}
Where $\mathbf{a}\cdot \mathbf{x}=\sum_{i=1}^3a_ix_i$, $\mathbf{a}\cdot \mathbf{y}=\sum_{i=1}^3a_iy_i$,
$T=(T_{ij})$.
It is desirable but seems not easy to optimize Eq. (38), here we only discuss some special cases of it:

(i) if $T=0$, then $D_{AB}^G(\rho ^{AB})=0$;

(ii) if $\mathbf{x}=\mathbf{y}=0$, that is $\rho ^A=I_A$ and $\rho ^B=I_B$, by the singular value decomposition of T, we get
 $D_{AB}^G(\rho ^{AB})=\frac 14[tr(TT^t)-\lambda _{\max }]$, with $%
\lambda _{\max }$ being the largest eigenvalue of $TT^t$;

(iii) if $T_{ij}=x_iy_j$, that is $\rho ^{AB}=\rho ^A\otimes \rho ^B$, then $%
D_{AB}^G(\rho ^{AB})=0$.

\section{Summary}

We introduced a geometric measure for the quantum discord defined over two-sided projective measurements, simplified the expression and provided a lower bound for this geometric measure. Some special quantum states were discussed as demonstrations of this geometric measure. We expect that this geometric measure may provide an new perspective and bring some conveniences for understanding and characterization of quantum discord over two-sided projective measurements.

 It has shown that $D_{AB}(\rho ^{AB})$ captures more correlation than $D_A(\rho ^{AB})$. At the end of this paper, it is interesting to point out the ordering of some different quantum correlation below
\begin{eqnarray}
\rho ^{AB}=\rho ^A\otimes \rho ^B
\Longrightarrow   D_{AB}(\rho ^{AB})=0
\Longrightarrow  D_A(\rho ^{AB})=0
\Longrightarrow  \rho ^{AB} \ is \ separable.  \nonumber
\end{eqnarray}

\section*{Acknowledgments}
This work was supported by National Natural Science Foundation of China
(Grant Nos. 10775101). The author thanks Qing Hou for helpful discussions.

\section*{References}
\bibliographystyle{model1-num-names}

\end{document}